\documentclass[12pt]{article}

\usepackage{amssymb}
\usepackage{color}
\usepackage{epsfig,amssymb,amsfonts,amsmath,graphicx,dsfont,cite,xfrac}
\usepackage{authblk}
\usepackage{braket,bbm}
\definecolor{mygray}{gray}{0.5}
\usepackage{subfigure}

\usepackage{cite}
\usepackage[colorlinks=true,linkcolor=blue,citecolor=red]{hyperref}

\newcommand{\be}{\begin{equation}}
\newcommand{\ee}{\end{equation}}
\newcommand{\bea}{\begin{eqnarray}}
\newcommand{\eea}{\end{eqnarray}}

\title{Factorization of the Quantum Fractional Oscillator}

\author[]{Fernando Olivar-Romero}
\author[]{Oscar Rosas-Ortiz}

\affil[]{\footnotesize Physics Department, Cinvestav, AP 14-740, 07000
M\'exico DF, Mexico}

\date{}
\begin{document}

\maketitle

\begin{abstract}
The applicability of the factorization method is extended  to the case of quantum fractional-differential Hamiltonians. In contrast with the conventional factorization, it is shown that the `factorization energy' is now a fractional-differential operator rather than a constant. As a first example, the energies and wave-functions of a fractional version of the quantum oscillator are determined. Interestingly, the energy eigenvalues are expressed as power-laws of the momentum in terms of the non-integer differential order of the eigenvalue equation.
\end{abstract}

\section{Introduction}

In fractional calculus the orders of integration and derivation are real numbers rather than natural ones. Thereby, besides the conventional derivatives of order $n \in \mathbb N$, one is able to evaluate derivatives of orders as arbitrary as $0.5$ or $\sqrt{2}$. The fractional derivatives are non-commutative and non-associative in general. Moreover, applied on a product of functions, they usually give rise to infinite series with derivatives and integrals of diverse non-integer orders (see e.g. \cite{Mil93,Kil06,Dua11}). Although such properties seem `counterintuitive', they find immediate application in describing systems with nonlocal properties of power-law type and are useful in different branches of physics and engineering (see e.g. \cite{Die10,Sab07,Hil00,Tar11}). Quite recently, the Feynman path integral approach was extended to the fractional case \cite{Las00a,Las00b}, the main idea was to replace the involved Brownian trajectories by L\'evy flights. As a consequence, a space-fractional version of the well known Schr\"odinger equation has come to light \cite{Las02}. Besides the cases of the quantum oscillator \cite{Las02} (see also \cite{Zab14}) and the hydrogen-like potential \cite{Las02}, the space-fractional Schr\"odinger equation has been solved for some piece-constant potentials \cite{Guo06,Zab14}, and has been extended to the time-fractional case \cite{Nab04}. Interesting discussions about the physical implementation of the quantum fractional oscillator can be found in \cite{Lon15,Zha15}. The physical realization of other fractional quantum systems is, as far as we know, an open problem.

With the present communication we extend the applicability of the factorization method \cite{Inf51,And84,Mie04} to the case of fractional-differential Hamiltonians. Without loss of generality, we pay attention to the one-dimensional oscillator and show that the `factorization energy', which is a constant in the conventional factorization, must be replaced by a fractional-differential operator in the extended fractional formulation. This last permits the application of algebraic methods to determine the energies and wave-functions of the quantum fractional oscillator. Remarkably, it is found that the energy eigenvalues are expressed as power-laws of the momentum in terms of the non-integer differential order of the eigenvalue equation. As indicated above, our method is not restricted to the oscillator-like interactions, it is also useful in  solving the fractional eigenvalue problem of other one-dimensional potentials.

\section{Problem and solution} 

For one-dimensional systems the time-independent, space-fractional Schr\"odinger equation \cite{Las02} is of the form
\begin{equation} 
\left[-{\cal D}_{\alpha}\hbar^{\alpha}\frac{d^{\alpha}}{dx^{\alpha}}+V(x)\right]\psi(x)=E\psi(x),
\end{equation}
where $\alpha$ is the L\'evy index ($1 < \alpha \leq 2$), $\frac{d^{\alpha}}{dx^{\alpha}}$ is the Riesz fractional derivative of order $\alpha$, and ${\cal D}_{\alpha}$ is a scale constant. The conventional Schr\"odinger equation is recovered if $\alpha =2$. From now on we use proper units such that ${\cal D}_{\alpha} \hbar^{\alpha}=1$. 

In \cite{Las02}, the fractional oscillator potential is defined as $\vert x \vert^{\beta}$, with $1< \beta \leq 2$. Here, for the sake of clarity, this parameter is fixed as $\beta=2$. The arbitrariness of $\beta$ makes no substantial difference in the method and can be retrieved in a further step. Therefore, the fractional-differential equation to be solved is given by
\begin{equation} 
H_{\alpha} \psi(x) \equiv \left[ -\frac{d^{\alpha}}{dx^{\alpha}}+x^{2}\right]\psi(x)=E\psi(x), \quad 1 < \alpha \leq 2.
\label{ES1}
\end{equation}
As this will be clear in the sequel, it is useful to express (\ref{ES1}) in momentum-representation
\begin{equation} 
\left[ \vert k \vert^{\alpha}-\frac{d^{2}}{dk^{2}}\right]\phi(k)=E\phi(k).
\label{ES2}
\end{equation}

\subsection{The fractional-factorization method} 
\label{gral}

We look for a pair of operators $A_{\alpha}$ and $B_{\alpha}$ such that 
\begin{equation} 
H_{\alpha}=B_{\alpha}A_{\alpha}+\epsilon_{\alpha},
\label{factor1}
\end{equation}
where $\epsilon_{\alpha}$ can be either a number (as in the conventional factorization) or a fractional-differential operator. The explicit form of the factors can be as general as in the conventional factorization whenever the identity (\ref{factor1}) is fulfilled. Here we present just one of their simpler expressions. Namely, 
\begin{equation} 
A_{\alpha}=\frac{d^{\alpha/2}}{dx^{\alpha/2}}+x \qquad \mbox{and} \qquad  B_{\alpha}=-\frac{d^{\alpha/2}}{dx^{\alpha/2}}+x
\end{equation}
give
\begin{equation} 
B_{\alpha}A_{\alpha} =-\frac{d^{\alpha}}{dx^{\alpha}}-\frac{\alpha}{2}\frac{d^{\alpha/2-1}}{dx^{\alpha/2-1}}+x^{2}.
\label{factor2}
\end{equation}
Comparing (\ref{factor2}) with (\ref{ES1}) we see that (\ref{factor1}) is fulfilled if
\begin{equation} 
\epsilon_{\alpha}=\frac{\alpha}{2}\frac{d^{\alpha/2-1}}{dx^{\alpha/2-1}}. 
\label{epsilon}
\end{equation}
That is, for $\alpha \neq 2$, the  factorization energy $\epsilon_{\alpha}$ is a fractional-differential operator of order $\tfrac{\alpha}{2}-1$ rather than a constant. Notice that $\alpha=2$ leads to the conventional result $\epsilon =1$, with $A_2=a$ and $B_2=a^{\dagger}$ the boson operators of the (mathematical) quantum oscillator $V_{osc}(x) =x^2$.

\subsection{Spectrum and wavefunctions} 

To identify the kernel of $A_{\alpha}$ we have to solve the fractional-differential equation
\begin{equation} 
A_{\alpha}\psi_{0}^{(\alpha)}(x)=\left[\frac{d^{\alpha/2}}{dx^{\alpha/2}}+x\right]\psi_{0}^{(\alpha)} (x)=0.
\label{ker1}
\end{equation}
Equivalently, one can solve this last equation in position-representation
\begin{equation} 
i\left[k^{\alpha/2} \mbox{sgn} (k) +\frac{d}{dk}\right]\phi_{0}^{(\alpha)} (k)=0.
\end{equation}
Substituting the solution
\begin{equation} 
\phi_{0}^{(\alpha)}(k)=\exp\left(-\frac{2|k|^{\alpha/2+1}}{\alpha+2}\right)
\label{phi0}
\end{equation}
into (\ref{ES2}) we obtain an expression for the related energy
\begin{equation} 
E_{0}(k, \alpha)=1 \left( \frac{\alpha}{2|k|^{1-\alpha/2}} \right).
\label{e0}
\end{equation}
Thus, the wave-function $\phi_0^{(\alpha)}(k)$ and energy $E_0(k,\alpha)$ have a power-law dependence on the momentum $k$ that is determined by the L\'evy index $\alpha$. In particular, the solutions for the ground state $E_0=1$ of the (mathematical) quantum oscillator are recovered if $\alpha=2$. Similar results have been reported for the  `Cauchy oscillator' by using the so called Strang splitting method in \cite{Zab14}. In our case, (\ref{phi0}) and (\ref{e0}) are just a consequence of the factorization (\ref{factor1}). Indeed, introducing (\ref{ker1}) in (\ref{factor1}) and using (\ref{ES1}), one obtains the equation
\begin{equation}
\left[ -\frac{d^{\alpha}}{dx^{\alpha}}+x^{2} -\frac{\alpha}{2}\frac{d^{\alpha/2-1}}{dx^{\alpha/2-1}}
\right]\psi_0^{(\alpha)}(x) =0.
\end{equation}
In momentum-representation this last acquires the form
\begin{equation}
\left[
\vert k \vert^{\alpha} - \frac{d^2}{dk^2} -\frac{\alpha}{2} \vert k \vert^{\alpha/2-1}
\right] \phi_0^{(\alpha)} =0.
\label{ke0}
\end{equation}
Clearly, the function (\ref{phi0}) is solution of (\ref{ke0}), so that our approach is self-consistent.

\begin{figure}[htp]
\centering
\subfigure[$\psi_0^{(\alpha)}(x)$]{\includegraphics[width=0.2\textwidth]{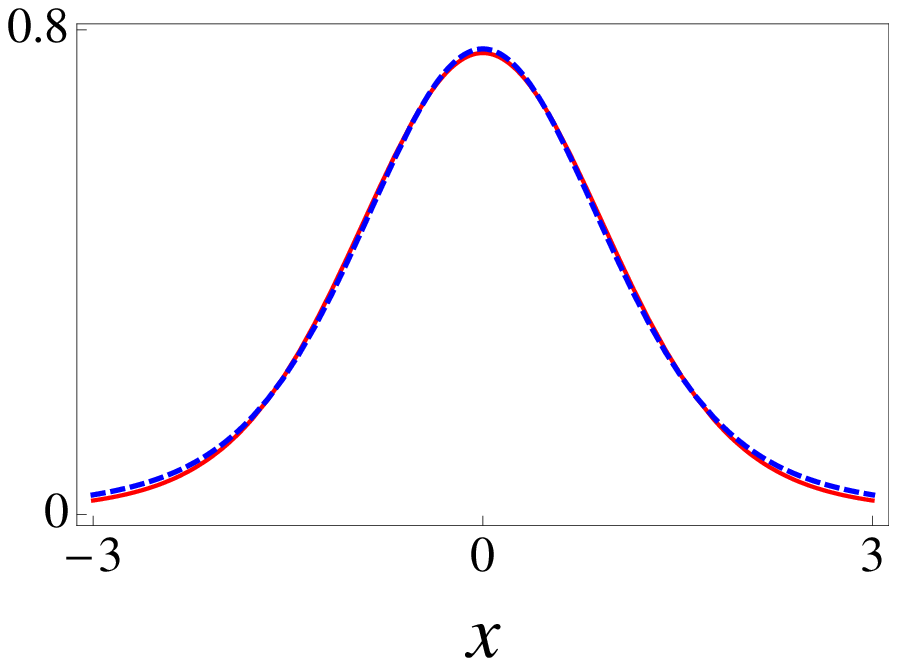}} 
\hspace{5mm}
\subfigure[$\psi_1^{(\alpha)}(x)$]{\includegraphics[width=0.2\textwidth]{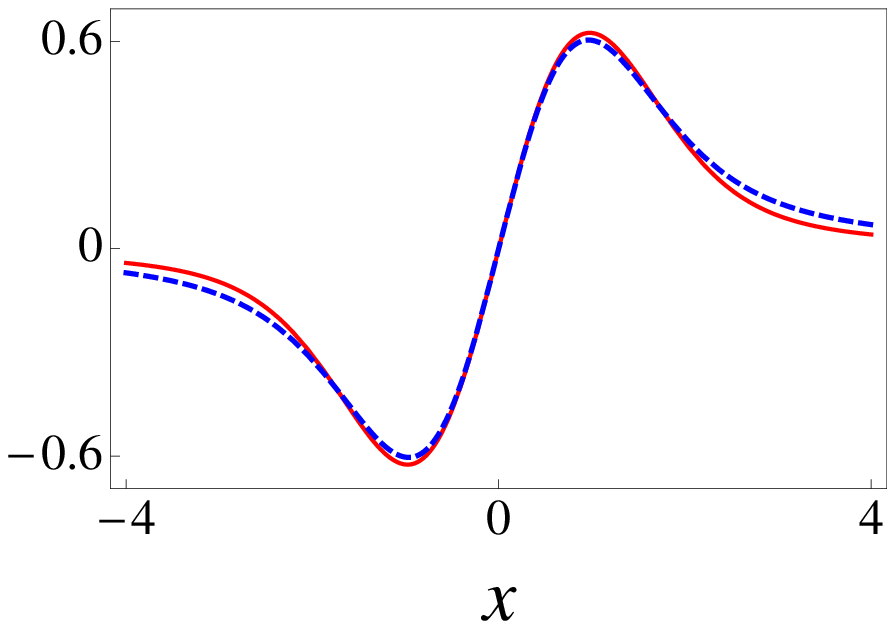}}
\hspace{5mm}
\subfigure[$\psi_2^{(\alpha)}(x)$]{\includegraphics[width=0.2\textwidth]{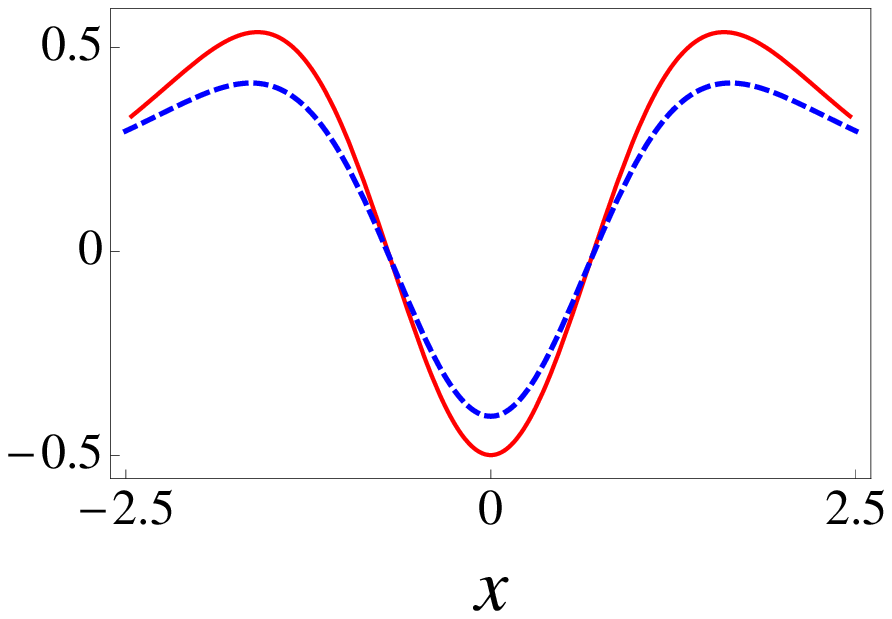}} 
\vskip1ex
\centering
\subfigure[$E_0(k,\alpha)$]{\includegraphics[width=0.2\textwidth]{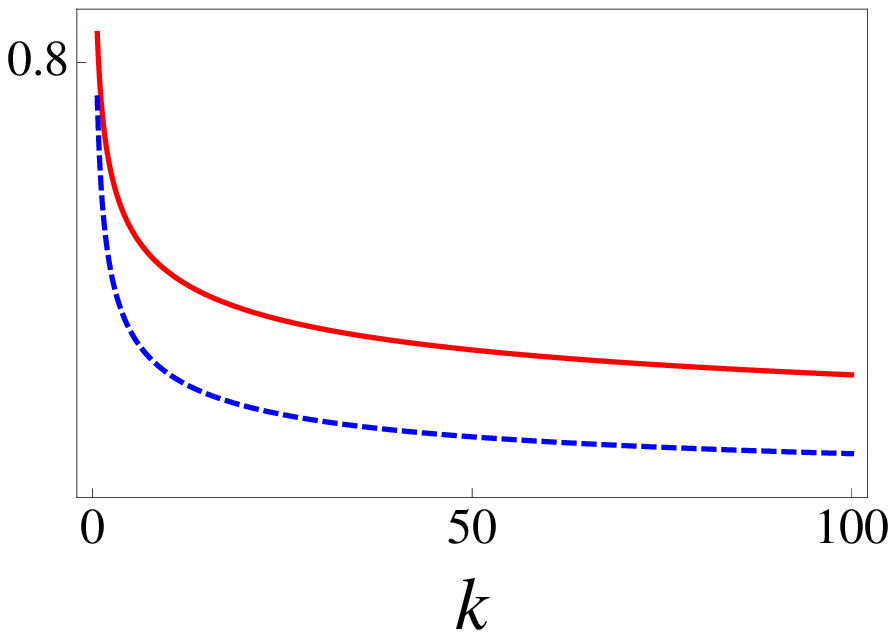}}
\hspace{5mm}
\subfigure[$E_1(k,\alpha)$]{\includegraphics[width=0.2\textwidth]{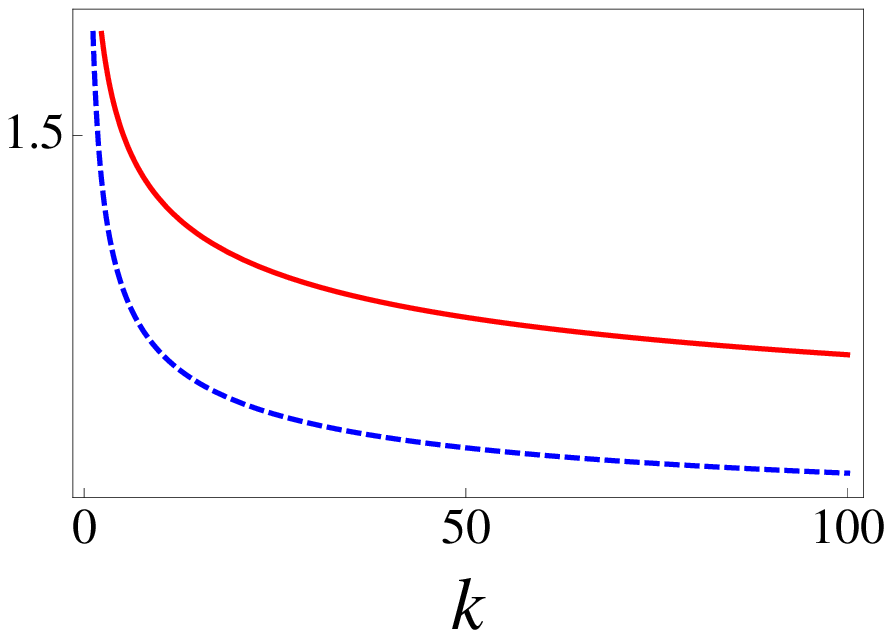}} 
\hspace{5mm}
\subfigure[$E_2(k,\alpha)$]{\includegraphics[width=0.2\textwidth]{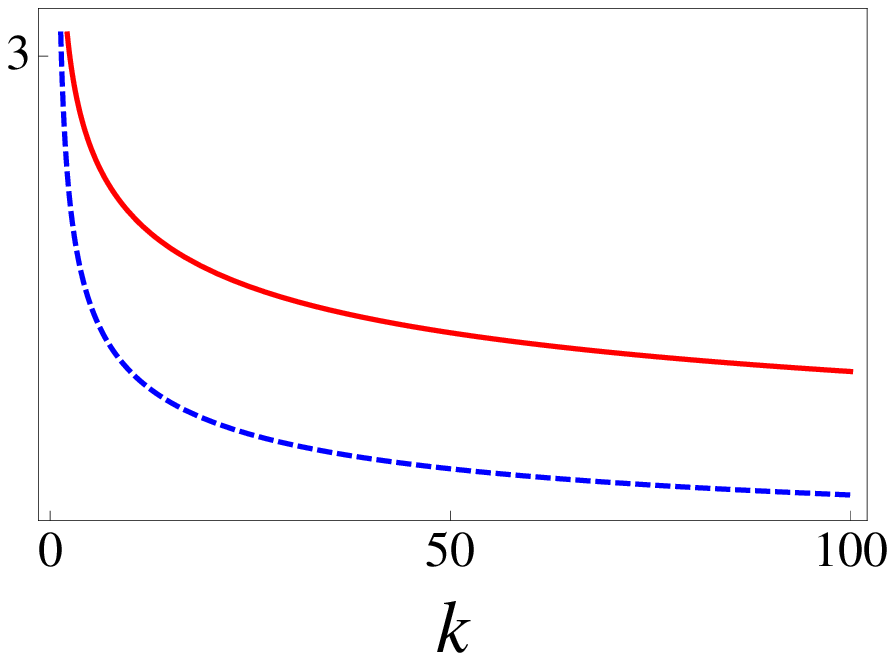}}

\caption{\footnotesize 
Wave-functions (a--c) and energies (d--f) of the first three states of the quantum fractional oscillator. In all cases the dashed-blue curve corresponds to $\alpha=1.2$ while the solid-red one is for $\alpha=1.5$. The solutions of the conventional quantum oscillator are recovered if $\alpha=2$.
}
\label{ground}
\end{figure}

The behaviour of the (numerically calculated) Fourier transform $\psi_0^{(\alpha)} (x)$ of (\ref{phi0}) and the energy (\ref{e0}) are depicted in Figure~\ref{ground}(a) and (d) for two different values of the L\'evy index. Note that the variations of $\phi_0^{(\alpha)}(x)$ with respect to the parameter $\alpha$ are almost negligible (the power-law dependence on $k$ is in the argument of an exponential function). The situation for the energy eigenvalue $E_0(k,\alpha)$ is quite different because the power-law dependence on $k$ is in the denominator of (\ref{e0}).

As in the conventional factorization, the proper application of the operators $A_{\alpha}$ and $B_{\alpha}$ give rise to the other solutions of the problem. For instance, to get the first excited state $\phi_1^{(\alpha)} (k)$ we apply $B_{\alpha}$ on $\phi_0^{(\alpha)}(k)$. After some calculations one gets
\begin{equation}  
\phi_{1}^{(\alpha)} (k)=-2i |k|^{\alpha/2}  \mbox{sgn}(k) \exp\left(-\frac{2|k|^{\alpha/2+1}}{\alpha+2}\right),\end{equation}
and using (\ref{ES2}) we arrive at the expression for the energy
\begin{equation} 
E_{1} (k, \alpha) = 3\left( \frac{\alpha}{ 2 |k|^{1-\alpha/2}} \right) -\frac{\alpha}{2 |k|^{2} }\left(\frac{\alpha}{2}-1\right).
\end{equation}
A similar procedure is used to construct the other excited sates. In particular, the second excited state 
\begin{equation} 
\phi_{2}^{(\alpha)} (k)=(\alpha|k|^{\alpha/2 -1}-4|k|^{\alpha}) \exp\left(-\frac{2|k|^{\alpha/2+1}}{\alpha+2}\right),   
\end{equation}
belongs to the energy
\begin{equation}
E_{2} (k, \alpha)= 5 \left( \frac{\alpha}{2 \vert k \vert^{1-\alpha/2}} \right) \left( \frac{\alpha-1 -2\vert k \vert^{3\alpha/2 -1}}{\frac{\alpha}{2} - 2 \vert k \vert^{\alpha/2+1}} \right)  + \frac{\alpha}{2 \vert k \vert^2} \left(\frac{\alpha}{2} -1 \right)  \left[\frac{2 - \frac{\alpha}{2} + \vert k \vert^{\alpha/2+1} }{ \frac{\alpha}{2}- 2 \vert k \vert^{\alpha/2+1}}
\right]
\end{equation}
The behaviour of these solutions can be appreciated in Figure~\ref{ground}. In all cases the conventional results of the quantum oscillator are recovered if $\alpha=2$. Note that the dependence of the wave-functions on the L\'evy index $\alpha$ starts to be significative for the excited states. 

\section{Discussion and further applications} 

With some further modifications, the method can be immediately applied to solve the eigenvalue problem of the fractional potential $\vert x \vert^{\beta}$, $1< \beta \leq 2$, discussed in \cite{Las02}. In such a case, the solutions depend on the parameters $\alpha$ and $\beta$, and behave quite similar to the ones presented in this communication. Noticeably, one can use  more general fractional-differential operators $A_{\delta}$ and $B_{\gamma}$, with L\'evy indices $\delta$ and $\gamma$, such that the identity $H_{\alpha} = A_{\delta} B_{\gamma} + \epsilon_{\delta,\gamma}$ is true. This last includes the case in which $\alpha=2$ from the very beginning. That is, the case in which we look for the fractional-factorization of the conventional Hamiltonian in quantum mechanics. As expected, even in this last case, the fractional-factorization program gives rise to new algebraic properties of the factorizing operators \cite{Oli14}. This last opens new possibilities for the trends in supersymmetric quantum mechanics as this last is based fundamentally in the factorization method. Other possible applications of the fractional-factorization presented in this paper include the paraxial theory of optical beam propagation \cite{Gut07a, Gut07b}. Further progress in these subjects will be presented elsewhere.

\section{Conclusions}

We have introduced an algebraic technique to solve the eigenvalue problem of the Laskin  time-independent, space-fractional Schr\"odinger equation \cite{Las02}. This is based on a modification of the well known factorization method and requires that the factorization energy, which is a constant in the conventional factorization, be expressed as a differential operator of non-integer order. Although we have specialized our discussion in the case of the (mathematical) oscillator potential $V_{osc} (x)=x^2$, the generalities of the method can be glimpsed from the expressions derived in Section~\ref{gral}. Thus, with the appropriate refinements, the fractional-factorization can be implemented in practically all the cases where the conventional factorization is known to work. Of special interest, we have shown that the energies of the quantum fractional oscillator studied here depend on the momentum in terms of a power-law that is determined by the non-integer order of the fractional eigenvalue equation. Some insights of such a behaviour can be obtained from the classical fractional version of the same potential  for which a dissipative-like term is included in the motion equation \cite{Oli14}. Definitive conclusions on the matter are in progress.

\subsection*{Acknowledgments}
The comments and suggestions by the anonymous referees are acknowledged. FOR acknowledges the funding received through a CONACyT Scholarship.


\end{document}